\documentclass[prd,aps,twocolumn,preprintnumbers,amsmath,amssymb,nofootinbib,superscriptaddress,notitlepage]{revtex4-1}

\usepackage{epsfig}
\usepackage[utf8]{inputenc}
\usepackage{color}
\usepackage{epstopdf}
\usepackage[colorlinks=true,
linkcolor=blue,
breaklinks=true,
urlcolor=blue,
citecolor=blue]{hyperref}

\newcommand{\nn}{\nonumber}
\newcommand{\be}{\begin{equation}}
\newcommand{\ee}{\end{equation}}
\newcommand{\bea}{\begin{eqnarray}}
\newcommand{\eea}{\end{eqnarray}}
\newcommand{\ba}{\begin{array}}
\newcommand{\ea}{\end{array}}
\newcommand{\bi}{\begin{itemize}}
\newcommand{\ei}{\end{itemize}}

\renewcommand{\vec}[1]{\mbox{\boldmath $#1 \!\!$ \unboldmath}}
\renewcommand{\slash}{/ \!\!\!\!\,}

\newcommand{\lf}{\left}
\newcommand{\rg}{\right}

\newcommand{\ucas}{\affiliation{University of Chinese Academy of Sciences, Beijing 100049, China}}

\newcommand{\ynu}{\affiliation{Department of Physics, Yunnan University, Kunming 650091, China}}

\newcommand{\csr}{\affiliation{Research Center for Hadron and CSR Physics, Lanzhou University and Institute of Modern Physics of CAS, Lanzhou 730000, China}}

\newcommand{\keylab}{\affiliation{State Key Laboratory of Heavy Ion Science and Technology, Institute of Modern Physics, Chinese Academy of Sciences, Lanzhou 730000, China}}

\begin{document}

\title{Spin density matrix  of baryon-antibaryon pairs in electron-positron annihilation \\ with $P$ and $CP$ violation including electron mass}

\author{Chun-Qiu Zhao}
\ynu
\keylab

\author{Xu Cao}\email{corresponding author: caoxu@impcas.ac.cn}
\keylab
\ucas
\csr

\author{Jian-Ping Dai}\email{corresponding author: daijianping@ynu.edu.cn}
\ynu

\date{\today}

\begin{abstract} 
  \rule{0ex}{3ex}
   
In the center-of-mass frame, the spin density matrix for octet baryon-antibaryon pairs produced in polarized electron-positron annihilation is unambiguously determined within the one-photon-exchange approximation. The formalism simultaneously incorporates parity ($P$) and combined parity-charge conjugation ($CP$) violation arising from the coupling of the intermediate virtual photon to the baryon pair. The finite electron mass is explicitly accounted for as a correction to these $P$ and $CP$ violating effects. Built on the spin projection operator, the derived full and compact spin density matrix provides a natural starting point for evaluating other quantities of physical interest --- particularly angular distributions of sequential decays and the quantum entanglement of hyperon pairs.

\end{abstract}

\maketitle

\section{Introduction}

Big Bang theory claims that the early Universe should have contained equal amounts of matter and antimatter~\cite{Gamow:1946eb}. As the Universe expanded and cooled, most of the matter and antimatter annihilated in pairs, leaving behind only a small residual amount of matter. In 1967, A.D.Sakharov proposed three necessary conditions for generating the observed matter–antimatter asymmetry, one of which is the violation of charge conjugation ($C$) symmetry, along with the combined violation of charge conjugation and parity ($CP$) symmetry~\cite{Sakharov:1967dj}. Experimentally, the violations of parity and charge conjugation  separately in weak interactions were confirmed between 1957 and 1958~\cite{Lee:1956qn,Wu:1957my,Goldhaber:1958nb}. $CP$ violation was first observed in the decay of neutral $K$ mesons in 1964~\cite{Christenson:1964fg}, followed by its observation in $B$ mesons in 2001~\cite{BaBar:2001pki,Belle:2001zzw}, and in $D$ mesons in 2019~\cite{LHCb:2019hro}. Most recently, in March 2025, the LHCb collaboration reported the first observation of $CP$ violation in baryon decays~\cite{LHCb:2025ray}. These observations are fully consistent with the magnitude of $CP$ violation predicted by the Standard Model, arising from the Kobayashi–Maskawa phase~\cite{Cabibbo:1963yz,Kobayashi:1973fv}. However, the $CP$-violating effects within the Standard Model are at least ten orders of magnitude too small to explain the observed matter–antimatter asymmetry in the Universe~\cite{Barr:1979ye}. Therefore, it is widely postulated that there exist external sources of $CP$ violation beyond the Standard Model. Consequently, tests of $CP$ symmetry have become a key avenue in the search for new physics~\cite{Dine:2003ax}.

The elastic electron-nucleon scattering is of its own fundamental importance for the measurement of electromagnetic form factors (EMFFs) of nucleons \cite{Hofstadter:1956qs}.
Its timelike correspondence, the baryon-antibaryon production via electron-positron annihilation $e^+ e^- \to Y \bar Y$, is connected to spacelike region by dispersion relation \cite{Lin:2021umk,Lin:2021umz,Lin:2021xrc}, and equally important for exploring baryon structure \cite{Denig:2012by,Pacetti:2014jai,Dai:2024lau}.
Some unusual phenomena are observed in the timelike region.
For instance, periodic oscillation structures of nucleon effective form factors \cite{Bianconi:2015owa} is a manifestation of light unflavored vector mesons \cite{deMelo:2008rj,Cao:2018kos,Cao:2024tvz,Dai:2023vsw,Cao:2021asd}, or final state interaction \cite{Qian:2022whn,Yang:2024iuc,Yang:2022qoy}, or both.
Another interesting fact is that the baryon and antibaryon are polarized and correlated because of gauge invariance even the initial electron and positron beams are not polarized.
The polarization and correlation of hyperon and antihyperon could be measured by analyzing the angular distribution of their decay products \cite{Lee:1956qn}.
This makes it feasible to simultaneously measure the hyperon and anti-hyperon decay parameters so precisely probe the violations of the combined parity and charge-conjugate $CP$ symmetry between them \cite{Faldt:2017kgy}.
The BESIII collaboration has extensively tested $CP$ symmetry exploiting hyperon pairs production ~\cite{BESIII:2018cnd,BESIII:2019cuv,BESIII:2021ypr,BESIII:2021cvv,BESIII:2022qax,BESIII:2022lsz,BESIII:2023drj,BESIII:2023sgt,BESIII:2023cvk,BESIII:2024nif,BESIII:2025jxt},
enabling the most precise examination to date of $CP$ violation in baryon decays~\cite{BESIII:2022qax} based on the world’s largest $J/\psi$ dataset~\cite{BESIII:2020nme,BESIII:2021cxx}.

It is proposed that the parity $P$ and $CP$ symmetries in the charmonium coupling to the baryon pairs \cite{He:1992ng,He:1993ar,He:2022jjc,Du:2024jfc} could be investigated in the same manner
in light of the tens of billions of $J/\psi$ events already collected by BESIII and expected larger data sets at future super tau-charm facility (STCF).
This scenario permits a measurement of the weak mixing angle and obtain the upper bounds on the electric dipole moments of the octet baryons at charmonium peak energies \cite{Du:2024jfc,Chen:2025rab}.
The complete angular distributions in the whole decay chain are written within helicity representation \cite{Zhang:2010zzo,Zhang:2009at,Fu:2023ose,Ovsiannikov:2025gcy} and used to investigate the hyperon electric dipole moment (EDM) arising from $CP$ violation by BESIII collaboration \cite{BESIII:2025vxm}.
In this paper spin density matrix  for baryon-antibaryon pairs in this process is calculated with the help of covariant spin projector operator.
Both transverse and longitudinal polarization of beams are considered.
Furthermore, we explicitly separate and identify the contributions arising from the finite electron mass.
The two–photon exchange (TPE) is not considered here, as it has already been fully accounted for in the literature \cite{Gakh:2005hh,Gakh:2005wa,Adamuscin:2007xn,Chen:2008hka}.
While these contributions are small, accounting for them is essential when searching for similarly subtle  $P$ and $CP$-violating effects to reach a sound conclusion.
The spin density matrix can be directly used to investigate the quantum entanglement of hyperon-anti-hyperon pairs produced in electron-positron annihilation \cite{Tornqvist:1980af,Qian:2020ini,Fabbrichesi:2024rec,Wu:2024mtj,Hong:2025drg,Zhang:2026nwm,Li:2026bkf}.

The paper is organized in the following way. Section~\ref{sectionunpol} presents the unpolarized cross section. Section~\ref{singlespinpol} discusses the single-spin polarization observables, and Section~\ref{doublespinpol} focuses on the double-spin polarization observables. By combing the results in previous two section, spin density matrix are given in an elaborated way in Section~\ref{correlationmatrix}. The conclusions are breafly summarized in Section~\ref{conclusion}.

\section{Unpolarized cross section} \label{sectionunpol}

We perform the calculation in the center-of-mass c.m. frame with the $z$-axis aligning along the three-momentum of the antibaryon. 
The amplitude of electron–positron annihilation into octet baryon-antibaryon incorporating $P$ and $CP$ violation at the tree level is \cite{He:2022jjc,Guo:2025bfn}:
\begin{equation}
\begin{split}
&\mathcal{M} = -\frac{e^2}{q^2} \bar{v}(k_2)\gamma_{\mu}u(k_1) \times \\
&\bar{u}(p_2) \biggl(F_1 \gamma^{\mu} + \frac{i F_2}{2M} \sigma^{\mu\nu} q_\nu + F_A \gamma^{\mu} \gamma_5 + \frac{H_T}{2M}\sigma^{\mu\nu} q_\nu \gamma_5 \biggr) v(p_1)\,, \qquad
\label{eq:matPCP}
\end{split}
\end{equation}
with $q = p_1 + p_2$ being the momentum transfer, and $M$ the baryon mass. 
The $F_1$ and $F_2$ are Dirac and Pauli form factors, respectively. The $F_A$ and $H_T$ represent the dimensionless form factors associated with $P$ and $CP$ violating strengths.
Note that in this paper dimensionless $H_T$ is used and differs from other work by a factor of $2M$.
All the form factors receive contributions from the exchange of one-photon including possible one-photon conversion to charmonium and $Z$-boson exchange $et al.$.
The form factor $H_T$ as a function of the squared momentum transfer $q^2=s=(p_1+ p_2)^2$ is related to the electric dipole moment (EDM) of the baryon in the $q^2 \to 0$ limit~\cite{He:1993ar,Du:2024jfc,Zhang:2009at,Zhang:2010zzo}. 
The form factor $F_A$ stems from the parity-violating decay of charmonium, which in the Standard Model receives contributions from $s$-channel $Z$-boson exchange and $t$-channel $W$-boson exchange~\cite{Du:2024jfc}.
Later on, Sachs magnetic ($G_M$) and charge ($G_E$) EMFFs are used, which are related to $F_1$ and $F_2$ as follows:
\bea
G_M=F_1+F_2 \,, G_E=F_1+\tau F_2 \,,
\eea
with $\tau={s}/{4M^2}$.
For convenience, Eq. (\ref{eq:matPCP}) is rewritten as:
\bea
\mathcal{M} = -\frac{e^2}{q^2} j_{\mu} \left( J_{EM}^{\mu}  + J_{A}^{\mu} + J_{T}^{\mu} \right) \,,
\eea
where :
\begin{equation}
\begin{aligned}
&j_{\mu} = \bar{v}(k_2) \gamma_{\mu} u(k_1) \,,\\
&J_{EM}^{\mu} = \bar{u}(p_2) \left( F_1 \gamma^{\mu} + \frac{i F_2}{2 M} \sigma^{\mu\nu} q_{\nu} \right) v(p_1) \,,\\
&J_{A}^{\mu} = F_A \bar{u}(p_2) \gamma^{\mu} \gamma_5 v(p_1),\\
&J_{T}^{\mu} = \frac{H_T}{2 M} \bar{u}(p_2) \sigma^{\mu\nu} q_{\nu} \gamma_5 v(p_1) \,.
\label{LHflux}
\end{aligned}
\end{equation}

After summing over the polarizations of the baryon–antibaryon pair and averaging over the polarizations of the electron and positron, the spin-averaged differential cross section is given by:
\bea
\frac{d\sigma}{d\Omega} 
&= \frac{\alpha^2 \beta}{4 q^6} L_{\mu\nu} \biggl(H^{\mu\nu}_{EM} + H^{\mu\nu}_{A} + H^{\mu\nu}_{T} + 2\mathfrak{Re}H^{\mu\nu}_{EMA} \nonumber\\
&\quad + 2\mathfrak{Re}H^{\mu\nu}_{EMT} + 2\mathfrak{Re}H^{\mu\nu}_{TA}\biggr) \,,
\label{eq:UPcross-section}
\eea
with:
\begin{equation}
\begin{aligned}
&L_{\mu\nu} = j_{\mu} j_{\nu}^* \,, \\
&H^{\mu\nu}_{EM} = J_{EM}^{\mu} (J_{EM}^{\nu})^* \,, &
&H^{\mu\nu}_{A} = J_{A}^{\mu} (J_{A}^{\nu})^* \,, \\
&H^{\mu\nu}_{T} = J_{T}^{\mu} (J_{T}^{\nu})^* \,, &
&H^{\mu\nu}_{EMA} = J_{EM}^{\mu} (J_{A}^{\nu})^* \,, \\
&H^{\mu\nu}_{EMT} = J_{EM}^{\mu} (J_{T}^{\nu})^* \,, &
&H^{\mu\nu}_{TA} = J_{T}^{\mu} (J_{A}^{\nu})^* \,.
\label{LHtensors}
\end{aligned}
\end{equation}

Since the $z$-axis of the coordinate system defined in the CMS is along the direction of the antibaryon momentum, the four-momenta of the initial and final state particles can be easily written down as follows:
\begin{equation}
\begin{aligned}
p_1=(E,0,0,|\vec{p}_1|) \: &;& \: k_1=(E,-|\vec{k}_1|\sin\theta,0,|\vec{k}_1|\cos\theta)  \\
p_2=(E,0,0,-|\vec{p}_1|)  \: &;& \: k_2=(E,|\vec{k}_1|\sin\theta,0,-|\vec{k}_1|\cos\theta)  
\label{CMSvectors}
\end{aligned}
\end{equation}
where: $|\vec{k}_1|=\sqrt{E^2-m^2}$, $|\vec{p}_1|=\sqrt{E^2-M^2}$, with $m$ being mass of the initial electron, and $\theta$ being the center-of-mass scattering angle.

The density matrix for both initial electron(or positron) and final baryon(or antibaryon) is given by~\cite{Cao:2024tvz,Gakh:2005hh,Gakh:2005wa,Adamuscin:2007xn,Buttimore:2006mq}:
\begin{equation}
\begin{aligned}
\rho = u(p)\bar{u}(p) = (\slash{p} + \mathfrak{m})\frac{1}{2}(1 - \gamma_5 \slash{s}),\\
\rho = v(p)\bar{v}(p) = (\slash{p} - \mathfrak{m})\frac{1}{2}(1 - \gamma_5 \slash{s}).
\label{densitymatrix}
\end{aligned}
\end{equation}
After a Lorentz boost of the polarization unit vector $\vec{\xi}$ in the rest frame, the polarization four-vector $s_\mu$ of a relativistic particles is obtained
\bea
s_0=\frac{1}{\mathfrak{m}}\vec{p}\cdot \vec{\xi} \: ; \: \vec{s}=\vec{\xi} + \frac{\vec{p}(\vec{p}\cdot \vec{\xi})}{\mathfrak{m}(\mathfrak{m}+E)} \,. \label{4pol}
\eea
with $\vec{p}$, $E$ and $\mathfrak{m}$ being the three-momentum, energy and mass of the particles, respectively. Then the longitudinal polarization four-vectors $s_\mu$ for electron and positron are
\begin{equation}
\begin{aligned}
s_{e}^{\mu}(L)&=P_{L}(\sqrt{\tau_e-1}, -\sqrt{\tau_e}\sin\theta, 0, \sqrt{\tau_e}\cos\theta)\,,\\
s_{p}^{\mu}(L)&=\bar{P}_{L}(\sqrt{\tau_e-1}, \sqrt{\tau_e}\sin\theta, 0, -\sqrt{\tau_e}\cos\theta) \,,
\label{eplongFV}
\end{aligned}
\end{equation}
with $\tau_e={s}/{4m^2}$ and $P_L$ (or $\bar{P}_L)$ being the longitudinally polarization degree of electron (or positron) beams. If positron and electron beams have the same polarization vector in the individual helicity frame \cite{Cao:2024tvz}, transverse polarization four-vector $s_\mu$ are:
\begin{equation}
\begin{aligned}
s_{e}^{\mu}(T)&=P_{T}(0, \cos\theta\sin\phi, \cos\phi, \sin\theta\sin\phi)\,,\\
s_{p}^{\mu}(T)&=\bar{P}_{T}(0, \cos\theta\sin\phi, \cos\phi, \sin\theta\sin\phi) \,,
\label{eptransFV}
\end{aligned}
\end{equation}
with $P_T$ (or $\bar{P}_T)$ being the transversely polarization degree of electron (or positron) beams, and $\phi$ being the azimuthal angle of the polarization. 
Note that the definition of $\phi$ differs by a factor of $\pi$ in some literature.
A more general definition of the polarization vector in the literature \cite{Zhang:2025oks} does not affect the main conclusions of this work.

The differential cross section of $e^+e^- \to Y\bar{Y}$ is
\bea
\frac{d\sigma}{d\Omega}= \frac{\beta}{64\pi^2 q^2}|\mathcal{M}|^2 \equiv \frac{\alpha^2\beta }{4q^2}D
\label{UPcrossec},
\eea
where a flux factor $\beta$ is defined as the ratio of velocities between the final baryon and the initial lepton, e.g. $\beta=\sqrt{(s-4M^2)/(s-4m^2)}$.
By substituting Eqs. \eqref{LHflux} and \eqref{LHtensors} into Eq.(\ref{eq:UPcross-section}), the scaled angular distribution $D$ is obtained if electron and positron beams are both longitudinally polarized, 
\begin{widetext}
\bea
D&=&I_{L}^-\biggl((1+\cos^2\theta)|G_M|^2+\frac{1}{\tau}\sin^2\theta|G_E|^2+(1-\frac{1}{\tau})(1+\cos^2\theta)|F_A|^2+(\tau-1)\sin^2\theta|H_T|^2\biggr)\nn\\
&&+\frac{I_{L}^+}{\tau_e}\biggl(\sin^2\theta|G_M|^2+\frac{1}{\tau}\cos^2\theta|G_E|^2+(1-\frac{1}{\tau})\sin^2\theta|F_A|^2+(\tau-1)\cos^2\theta|H_T|^2\biggr)\nn\\
&&-4 \Delta L \sqrt{1-\frac{1}{\tau}}\cos\theta\mathfrak{Re}[G_M F_A^*]\,,
\label{eq:EPlongunpolDCPV}
\eea
\end{widetext}
with $I_{L}^{\pm} = 1 \pm P_L\bar{P}_L$ and $\Delta L = P_L-\bar{P}_L$.
Thus the interference of magnetic and $P$ violating form factor induce a $\cos\theta$ dependence of angular distribution if beams are of longitudinally polarization.
Note that real part of the interference of electromagnetic form factor and TPE also exhibits a characteristic odd dependence on $\cos\theta$, even for unpolarized beams \cite{Adamuscin:2007xn}.

If electron and positron beams are both transversely polarized, $D$ is written as
\begin{widetext}
\bea
D&=&(1+\cos^2\theta)|G_M|^2+\frac{1}{\tau}\sin^2\theta|G_E|^2+(1-\frac{1}{\tau})(1+\cos^2\theta)|F_A|^2+(\tau-1)\sin^2\theta|H_T|^2\nn\\
&&-P_T^2\sin^2\theta\cos2\phi\biggl(|G_M|^2-\frac{1}{\tau}|G_E|^2+(1-\frac{1}{\tau})|F_A|^2-(\tau-1)|H_T|^2\biggr)\nn\\
&&+\frac{I_{T}^{+}}{\tau_e}\biggl(\sin^2\theta|G_M|^2+\frac{1}{\tau}\cos^2\theta|G_E|^2+(1-\frac{1}{\tau})\sin^2\theta|F_A|^2+(\tau-1)\cos^2\theta|H_T|^2\biggr)\nn\\
&&-\frac{4\Sigma_T}{\sqrt{\tau_e}}\sqrt{1-\frac{1}{\tau}}\sin\theta\sin\phi\mathfrak{Re}[G_MF_A^*]\,,
\label{eq:EPtranunpolDCPV}
\eea
\end{widetext}
with $P_T^2 = P_T\bar{P}_T$ , $I_{T}^{+} = 1+P_T^2$ and $\Sigma_T = P_T+\bar{P}_T$. 
Thus, the effect of beam transverse polarization on the angular distribution is characterized by a dependence on the azimuthal angle $\phi$.
Particularly for the electron mass correction the real part of the interference of magnetic and $P$ violating form factor induce a $\sin\theta \sin\phi$ dependence of angular distribution when at least one of the beams is polarized.
The result with the massless electron limit and $P$ and $CP$ conservation aligns with previous findings in the helicity formalism \cite{Cao:2024tvz}.

When the beam polarization degree is zero ($I_{L}^{\pm} \to 1$, $\Delta L \to 0$, or $I_{T}^{+} \to 1$, $\Sigma_T \to 0$), $D$ reduces to the unpolarized-beam case~\cite{Buttimore:2006mq}.
The correction from finite electron mass to differential cross section (and spin polarization observables in later sections) is in the order of $1/\tau_e$. 
For a reference, one has $1/\tau_e \simeq 10^{-7}$ at the $J/\psi$ peak.




\section{Single-spin polarization observables} \label{singlespinpol}

According to Eq.~\eqref{4pol}, the polarization vectors for the baryon and antibaryon in the c.m. frame are explicitly given as:
\bea
s_x^\mu(B) &=& s_x^\mu(\bar{B})=(0, 1, 0, 0) \,,\nn \\
s_y^\mu(B) &=& s_y^\mu(\bar{B})=(0, 0, 1, 0) \,, \nn\\
s_z^\mu(B) &=& (-\sqrt{\tau-1}, 0, 0, \sqrt{\tau}) \,,\nn\\
s_z^\mu(\bar{B}) &=& (\sqrt{\tau-1}, 0, 0, \sqrt{\tau})\,.\nn
\eea
By inserting the spin projector operator above into Eq. \eqref{LHflux}, the single-spin polarization of the baryon and antibaryon in the case of longitudinally polarization beam could be directly calculated as,
\begin{widetext}
\bea
P_y(\bar{Y})&=&\frac{1}{D\sqrt{\tau}}(I_{L}^{-}-\frac{I_{L}^{+}}{\tau_e})\sin2\theta 
\biggl( \mathfrak{Im}[G_MG_E^*]-(\tau-1)\mathfrak{Re}[H_TF_A^*]\biggr) 
+\Delta L\frac{2\sqrt{\tau-1}}{D\tau}\sin\theta \biggl(\mathfrak{Im}[G_EF_A^*]+\tau\mathfrak{Re}[G_MH_T^*] \biggl)\,, \quad
\label{eq:EPlongpolPyCPV} \\
P_x(\bar{Y})&=&\frac{\sqrt{\tau-1}}{D\tau}(I_{L}^{-}-\frac{I_{L}^{+}}{\tau_e})\sin2\theta 
\biggl(\mathfrak{Re}[G_EF_A^*]+\tau\mathfrak{Im}[G_MH_T^*]\biggl) +\Delta L\frac{2\sin\theta}{D\sqrt{\tau}}\biggl((\tau-1)\mathfrak{Im}[H_TF_A^*]-\mathfrak{Re}[G_MG_E^*]\biggl)\,,
\label{eq:EPlongpolPxCPV} \\
P_z(\bar{Y})&=&I_{L}^{-}\frac{2\sqrt{\tau-1}}{D\sqrt{\tau}} \biggl(\sin^2\theta\mathfrak{Im}[G_EH_T^*]-(1+\cos^2\theta)\mathfrak{Re}[G_MF_A^*]\biggl) 
-\frac{I_{L}^{+}}{\tau_e}\frac{2\sqrt{\tau-1}}{D\sqrt{\tau}}\biggl(\sin^2\theta\mathfrak{Re}[G_MF_A^*]-\cos^2\theta\mathfrak{Im}[G_EH_T^*]\biggl) \nn \\
&&
+\Delta L\frac{2\cos\theta}{D\tau}\biggl(\tau|G_M|^2+(\tau-1)|F_A|^2\biggl)\,,
\label{eq:EPlongpolPzCPV}
\eea
\end{widetext}
During the calculation, the non-zero contraction of antisymmetric tensors with momenta in Appendix \ref{apdx:asy} are used.
It is observed that longitudinal polarization of the beam introduces an additional $\cos\theta$ dependence in $P_z$ and a $\sin\theta$ dependence in $P_x$.
A $\sin\theta$ dependence in $P_y$, however, arises only if $P$ or $CP$ symmetry is violated.
Note that the interference of EMFFs and TPE also exhibits a characteristic odd-power terms in $\cos\theta$ for those single-spin polarization observables \cite{Adamuscin:2007xn}.

In the case of transverse polarization beam:
\begin{widetext}
\bea
P_y(\bar{Y})&=&(1-\frac{I_{T}^{+}}{\tau_e})\frac{\sin2\theta}{D\sqrt{\tau}}\biggl( \mathfrak{Im}[G_MG_E^*]-(\tau-1)\mathfrak{Re}[H_TF_A^*]\biggl) 
+P_T^2\frac{\sin2\theta\cos2\phi}{D\sqrt{\tau}}\biggl( \mathfrak{Im}[G_MG_E^*]-(\tau-1)\mathfrak{Re}[H_TF_A^*]\biggl) \nn \\
&&-P_T^2\frac{2\sqrt{\tau-1}}{D\tau}\sin\theta\sin2\phi\biggl(\mathfrak{Re}[G_EF_A^*]+\tau\mathfrak{Im}[G_MH_T^*]\biggl)
+\frac{\Sigma_T}{\sqrt{\tau_e}}\frac{2\cos\phi}{D\sqrt{\tau}}\biggl(\mathfrak{Re}[G_MG_E^*]-(\tau-1)\mathfrak{Im}[H_TF_A^*]\biggl) \nn \\
&&-\frac{\Sigma_T}{\sqrt{\tau_e}}\frac{2\sqrt{\tau-1}}{D\tau}\cos\theta\sin\phi\biggl(\mathfrak{Im}[G_EF_A^*]+\tau\mathfrak{Re}[G_MH_T^*]\biggl)\,,
\label{eq:EPtranpolPyCPV}\\
P_x(\bar{Y})&=&(1-\frac{I_{T}^{+}}{\tau_e})\frac{\sqrt{\tau-1}}{D\tau}\sin2\theta\biggl(\mathfrak{Re}[G_EF_A^*]+\tau\mathfrak{Im}[G_MH_T^*]\biggl)
+P_T^2\frac{2\sin\theta\sin2\phi}{D\sqrt{\tau}}\biggl(\mathfrak{Im}[G_MG_E^*]-(\tau-1)\mathfrak{Re}[H_TF_A^*]\biggl) \nn \\
&&+P_T^2\frac{2\sqrt{\tau-1}}{D\tau}\sin2\theta\cos2\phi\biggl(\mathfrak{Re}[G_EF_A^*]+\tau\mathfrak{Im}[G_MH_T^*]\biggl)
+\frac{\Sigma_T}{\sqrt{\tau_e}}\frac{2\cos\theta\sin\phi}{D\sqrt{\tau}}\biggl(\mathfrak{Re}[G_MG_E^*]-(\tau-1)\mathfrak{Im}[H_TF_A^*]\biggl) \nn\\
&&+\frac{\Sigma_T}{\sqrt{\tau_e}}\frac{2\sqrt{\tau-1}}{D\tau}\cos\phi\biggl(\mathfrak{Im}[G_EF_A^*]+\tau\mathfrak{Re}[G_MH_T^*]\biggl)\,,
\label{eq:EPtranpolPxCPV} \\
P_z(\bar{Y})&=&-\frac{2\sqrt{\tau-1}}{D\sqrt{\tau}}\biggl((1+\cos^2\theta+\frac{I_{T}^{+}}{\tau_e}\sin^2\theta)\mathfrak{Re}[G_MF_A^*]-(\sin^2\theta+\frac{I_{T}^{+}}{\tau_e}\cos^2\theta)\mathfrak{Im}[G_EH_T^*]\biggl) \nn \\
&&+P_T^2\frac{2\sqrt{\tau-1}}{D\sqrt{\tau}}\sin^2\theta\cos2\phi\biggl(\mathfrak{Re}[G_MF_A^*]+\tau\mathfrak{Im}[G_EH_T^*]\biggl)+\frac{\Sigma_T}{\sqrt{\tau_e}}\frac{2\sin\theta\sin\phi}{D\tau}\biggl(\tau|G_M|^2+(\tau-1)|F_A|^2\biggl)\,,
\label{eq:EPtranpolPzCPV}
\eea
\end{widetext}
which is consistent with the results obtained by neglecting the electron mass and $P$ and $CP$ violation within the helicity formalism \cite{Cao:2024tvz}.
The electron mass correction introduces a $\sin\phi$ or $\cos\phi$ dependence to the single-spin polarization observables when at least one of the beams is polarized, e.g. those terms propotional to $\Sigma_T /\tau_e$, confronting with the normal $\sin2\phi$ or $\cos2\phi$ modulation resulting from simultaneously transverse polarization of the beams.

For both cases of unpolarized and polarized beams $P_y(Y)=P_y(\bar{Y}) [H_T \to -H_T]$, $P_x(Y)=P_x(\bar{Y}) [H_T \to -H_T]$, $P_z(Y)=P_z(\bar{Y}) [H_T \to -H_T]$ hold in the convention of helicity frame here.
For the case of unpolarized beams, the real part of interference of $P$ and $CP$ violation terms contributes to $P_y$, whereas 
the interference between electro- (or magnetic-) and $CP$ violation terms contributes to $P_z$ (or $P_x$), thereby inducing a difference in single-spin polarizations for baryon and antibaryon.
In the case of polarized beams, The $P$ and $CP$ violation separately induce non-zero single-spin polarization observables through different interference manners with EMFFs.
Interestingly, the $\Delta L$ terms originate from the same interference pattern as one of the $\Sigma_T/\tau_e$ terms, for instance $\Delta L \sin\theta$ and $\Sigma_T/\tau_e \cos\theta \sin\phi$ terms in $P_y$.

Above calculations clearly confirm that the final baryon and antibaryon are of transverse polarization in the $y$ direction even the initial electron is unpolarized in the case of $P$ and $CP$ conservation \cite{Faldt:2017kgy}.
A measurement of $P_y$ allows for determine the imaginary part of interference between electro- and magnetic form factors.
The electron mass correction introduces a new angular distribution in $P_z$ in Eq. \eqref{eq:EPlongpolPzCPV} for longitudinal beams, and contributes to all single-spin polarization observables for transverse beams.

\section{Double-spin polarization observables}\label{doublespinpol}

The double-spin polarization observables represent spin correlation between the final baryon and antibaryon. 
By inserting the spin projector operator of baryon and antibaryon defined above simultaneously into Eq. \eqref{LHflux}, the double-spin polarization of the baryon and antibaryon are given as
\begin{widetext}
\bea
C_{xx}&=&\frac{I_{L}^{-}}{D\tau}\sin^2\theta\biggl(\tau|G_M|^2+|G_E|^2-(\tau-1)|F_A|^2-\tau(\tau-1)|H_T|^2\biggr)\nn\\
&&-\frac{I_{L}^{+}}{\tau_e}\frac{1}{D\tau}\biggl(\tau\sin^2\theta|G_M|^2-\cos^2\theta|G_E|^2-(\tau-1)\sin^2\theta|F_A|^2+\tau(\tau-1)\cos^2\theta|H_T|^2\biggr)\,,
\label{eq:EPlongpolPxxCPV}\\
C_{yy}&=&\frac{I_{L}^{-}}{D\tau}\sin^2\theta\biggl(-\tau|G_M|^2+|G_E|^2+(\tau-1)|F_A|^2-\tau(\tau-1)|H_T|^2\biggr)\nn\\
&&+\frac{I_{L}^{+}}{\tau_e}\frac{1}{D\tau}\biggl(\tau\sin^2\theta|G_M|^2+\cos^2\theta|G_E|^2-(\tau-1)\sin^2\theta|F_A|^2-\tau(\tau-1)\cos^2\theta|H_T|^2\biggr) \,,
\label{eq:EPlongpolPyyCPV}\\
C_{zz}&=&\frac{I_{L}^{-}}{D\tau}\biggl(\tau(1+\cos^2\theta)|G_M|^2-\sin^2\theta|G_E|^2+(\tau-1)(1+\cos^2\theta)|F_A|^2-\tau(\tau-1)\sin^2\theta|H_T|^2\biggr)\nn\\
&&+\frac{I_{L}^{+}}{\tau_e}\frac{1}{D\tau}\biggl(\tau\sin^2\theta|G_M|^2-\cos^2\theta|G_E|^2+(\tau-1)\sin^2\theta|F_A|^2-\tau(\tau-1)\cos^2\theta|H_T|^2\biggr)\nn\\
&&-\frac{4 \Delta L}{D}\sqrt{1-\frac{1}{\tau}}\cos\theta\mathfrak{Re}[G_M F_A^*]\,,
\label{eq:EPlongpolPzzCPV}\\
C_{xy}&=&I_{L}^{-}\frac{2\sqrt{\tau-1}\sin^2\theta}{D\sqrt{\tau}}\biggl(-\mathfrak{Im}[G_M F_A^*]+\mathfrak{Re}[G_EH_T^*]\biggr)+\frac{I_{L}^{+}}{\tau_e}\frac{2\sqrt{\tau-1}}{D\sqrt{\tau}}\biggl(\sin^2\theta\mathfrak{Im}[G_M F_A^*]+\cos^2\theta\mathfrak{Re}[G_EH_T^*]\biggr)\,,
\label{eq:EPlongpolPxyCPV}\\
C_{xz}&=&\frac{1}{D\sqrt{\tau}}(I_{L}^{-}-\frac{I_{L}^{+}}{\tau_e})\sin2\theta\biggl(-\mathfrak{Re}[G_MG_E^*]+(\tau-1)\mathfrak{Im}[H_TF_A^*]\biggr) \nn\\
&&+\Delta L\frac{2\sqrt{\tau-1}}{D\tau}\sin\theta\biggl(\mathfrak{Re}[G_EF_A^*]+\tau\mathfrak{Im}[G_MH_T^*]\biggr) \,,
\label{eq:EPlongpolPxzCPV}\\
C_{yz}&=&\frac{\sqrt{\tau-1}\sin2\theta}{D\tau}(I_{L}^{-}-\frac{I_{L}^{+}}{\tau_e})\biggl(\mathfrak{Im}[G_EF_A^*]+\tau\mathfrak{Re}[G_MH_T^*]\biggr)+\Delta L\frac{2\sin\theta}{D\sqrt{\tau}}\biggl(\mathfrak{Im}[G_MG_E^*]-(\tau-1)\mathfrak{Re}[H_TF_A^*]\biggr) \,,
\label{eq:EPlongpolPyzCPV}
\eea
\end{widetext}
if electron and positron beams are both longitudinally
polarized.
The longitudinal polarization beam introduce a new $\cos\theta$ distribution in $C_{zz}$ in Eq. \eqref{eq:EPlongpolPzzCPV} when $P$ is violated, and a new $\sin\theta$ dependence in $C_{xz}$ and $C_{yz}$ when $P$ or $CP$ is violated.
The $C_{xx}$, $C_{yy}$ and $C_{zz}$ are characterized by $\cos^2\theta$ modulation originated from squared moduli of various form factors, but with an exception of a $\cos\theta$ dependence in $C_{zz}$, the same term in angular distribution $D$ in Eq. \eqref{eq:EPlongunpolDCPV}.
The interference between different form factors gives rise to non-zero non-diagonal correlations.

In the case of transverse polarization beam, double-spin polarization observables are
\begin{widetext}
\bea
C_{xx}&=&\frac{\sin^2\theta}{D\tau}\biggl(\tau|G_M|^2+|G_E|^2-(\tau-1)|F_A|^2-\tau(\tau-1)|H_T|^2\biggr)-P_T^2\frac{\cos2\phi}{D\tau}\biggl(\tau(1+\cos^2\theta)|G_M|^2-\sin^2\theta|G_E|^2\biggr)\nn\\
&&+P_T^2\frac{\tau-1}{D\tau}\cos2\phi\biggl((1+\cos^2\theta)|F_A|^2-\tau\sin^2\theta|H_T|^2\biggr)+P_T^2\frac{4\sqrt{\tau-1}}{D\sqrt{\tau}}\cos\theta\sin2\phi\mathfrak{Im}[G_MF_A^*]\nn\\
&&-\frac{I_{T}^{+}}{\tau_e}\frac{1}{D\tau}\biggl(\tau\sin^2\theta|G_M|^2-\cos^2\theta|G_E|^2\biggr)+\frac{I_{T}^{+}}{\tau_e}\frac{\tau-1}{D\tau}\biggl(\sin^2\theta|F_A|^2-\tau\cos^2\theta|H_T|^2\biggr)\,,
\label{eq:EPtranpolPxxCPV}\\
C_{yy}&=&-\frac{\sin^2\theta}{D\tau}\biggl(\tau|G_M|^2-|G_E|^2-(\tau-1)(|F_A|^2-\tau|H_T|^2)\biggr)+P_T^2\frac{\cos2\phi}{D\tau}\biggl(\tau(1+\cos^2\theta)|G_M|^2+\sin^2\theta|G_E|^2\biggr)\nn\\
&&-P_T^2\frac{\tau-1}{D\tau}\cos2\phi\biggl((1+\cos^2\theta)|F_A|^2+\tau\sin^2\theta|H_T|^2\biggr)-P_T^2\frac{4\sqrt{\tau-1}}{D\sqrt{\tau}}\cos\theta\sin2\phi\mathfrak{Im}[G_MF_A^*]\nn\\
&&+\frac{I_{T}^{+}}{\tau_e}\frac{1}{D\tau}\biggl(\tau\sin^2\theta|G_M|^2+\cos^2\theta|G_E|^2\biggr)-\frac{I_{T}^{+}}{\tau_e}\frac{\tau-1}{D\tau}\biggl(\sin^2\theta|F_A|^2+\tau\cos^2\theta|H_T|^2\biggr)\,,
\label{eq:EPtranpolPyyCPV}\\
C_{zz}&=&\frac{1}{D\tau}\biggl(\tau(1+\cos^2\theta)|G_M|^2-\sin^2\theta|G_E|^2+(\tau-1)(1+\cos^2\theta)|F_A|^2-\tau(\tau-1)\sin^2\theta|H_T|^2\biggl)\nn\\
&&-P_T^2\frac{\sin^2\theta\cos2\phi}{D\tau}\biggl(\tau|G_M|^2+|G_E|^2+(\tau-1)(|F_A|^2+\tau|H_T|^2)\biggr)+\frac{I_{T}^{+}}{\tau_e}\frac{1}{D\tau}\biggl(\tau\sin^2\theta|G_M|^2-\cos^2\theta|G_E|^2\biggr)\nn\\
&&+\frac{I_{T}^{+}}{\tau_e}\frac{\tau-1}{D\tau}\biggl(\sin^2\theta|F_A|^2-\tau\cos^2\theta|H_T|^2\biggr)-\frac{\Sigma_T}{\sqrt{\tau_e}}\frac{4\sqrt{\tau-1}}{D\sqrt{\tau}}\sin\theta\sin\phi\mathfrak{Re}[G_M F_A^*]\,,
\label{eq:EPtranpolPzzCPV}\\
C_{xy}&=&-\frac{2\sqrt{\tau-1}}{D\sqrt{\tau}}\sin^2\theta\biggl(\mathfrak{Im}[G_M F_A^*]-\mathfrak{Re}[G_EH_T^*]\biggr)+P_T^2\frac{2\sqrt{\tau-1}}{D\sqrt{\tau}}\cos2\phi\biggl((1+\cos^2\theta)\mathfrak{Im}[G_M F_A^*]+\sin^2\theta\mathfrak{Re}[G_EH_T^*]\biggr)\nn\\
&&+P_T^2\frac{2\cos\theta\sin2\phi}{D\tau}\biggl(\tau|G_M|^2-(\tau-1)|F_A|^2\biggr)+\frac{I_{T}^{+}}{\tau_e}\frac{2\sqrt{\tau-1}}{D\sqrt{\tau}}\biggl(\sin^2\theta\mathfrak{Im}[G_M F_A^*]+\cos^2\theta\mathfrak{Re}[G_EH_T^*]\biggr)\,,
\label{eq:EPtranpolPxyCPV}\\
C_{xz}&=&-(1-\frac{I_{T}^{+}}{\tau_e})\frac{\sin2\theta}{D\sqrt{\tau}}\biggl(\mathfrak{Re}[G_MG_E^*]-(\tau-1)\mathfrak{Im}[H_TF_A^*]\biggr)-P_T^2\frac{\sin2\theta\cos2\phi}{D\sqrt{\tau}}\biggl(\mathfrak{Re}[G_MG_E^*]-(\tau-1)\mathfrak{Im}[H_TF_A^*]\biggr)\nn\\
&&+P_T^2\frac{2\sqrt{\tau-1}}{D\tau}\sin\theta\sin2\phi\biggl(\mathfrak{Im}[G_EF_A^*]+\tau\mathfrak{Re}[G_MH_T^*]\biggr)+\frac{\Sigma_T}{\sqrt{\tau_e}}\frac{2\cos\phi}{D\sqrt{\tau}}\biggl(\mathfrak{Im}[G_MG_E^*]-(\tau-1)\mathfrak{Re}[H_TF_A^*]\biggr)\nn\\
&&-\frac{\Sigma_T}{\sqrt{\tau_e}}\frac{2\sqrt{\tau-1}}{D\tau}\cos\theta\sin\phi\biggl(\mathfrak{Re}[G_EF_A^*]+\tau\mathfrak{Im}[G_MH_T^*]\biggr)\,,
\label{eq:EPtranpolPxzCPV}\\
C_{yz}&=&(1-\frac{I_{T}^{+}}{\tau_e})\frac{\sqrt{\tau-1}}{D\tau}\sin2\theta\biggl(\mathfrak{Im}[G_EF_A^*]+\tau\mathfrak{Re}[G_MH_T^*]\biggr)+P_T^2\frac{2\sin\theta\sin2\phi}{D\sqrt{\tau}}\biggl(\mathfrak{Re}[G_MG_E^*]-(\tau-1)\mathfrak{Im}[H_TF_A^*]\biggr)\nn\\
&&+P_T^2\frac{\sqrt{\tau-1}}{D\tau}\sin2\theta\cos2\phi\biggl(\mathfrak{Im}[G_EF_A^*]+\tau\mathfrak{Re}[G_MH_T^*]\biggr)-\frac{\Sigma_T}{\sqrt{\tau_e}}\frac{2\cos\theta\sin\phi}{D\sqrt{\tau}}\biggl(\mathfrak{Im}[G_MG_E^*]-(\tau-1)\mathfrak{Re}[H_TF_A^*]\biggr)\nn\\
&&-\frac{\Sigma_T}{\sqrt{\tau_e}}\frac{2\sqrt{\tau-1}}{D\tau}\cos\phi\biggl(\mathfrak{Re}[G_EF_A^*]+\tau\mathfrak{Im}[G_MH_T^*]\biggr)\,,
\label{eq:EPtranpolPyzCPV}
\eea
\end{widetext}
which agrees with the results derived in the helicity formalism when the electron mass and $P$ and $CP$ violation is neglected \cite{Cao:2024tvz}.
The $C_{xx}$, $C_{yy}$ and $C_{zz}$ are characterized by $\cos^2\theta$ modulation originated from squared moduli of various form factors and a $\cos\theta$ or  $\sin\theta$ dependence, induced by
the interference of magnetic and $P$ violating form factor when at least one of the beams is polarized.
The $\cos\theta$ term in $C_{xx}$ is opposite to that in $C_{yy}$, and $\sin\theta$ term in $C_{zz}$ is exactly the same one in $D$ in Eq. \eqref{eq:EPtranunpolDCPV}, which is a new angular distribution introduced by electron mass correction.

For both cases of unpolarized and polarized beams $C_{yx}=C_{xy} [H_T \to -H_T]$, $C_{zx}=C_{xz} [H_T \to -H_T]$, $C_{zy}=C_{yz} [H_T \to -H_T]$ hold.
The non-diagonal correlations receive contribution not only from the interference between different form factors as the longitudinal beams case, but also from squared moduli of magnetic and $P$ violating form factors as encapsulated in $C_{xy}$ of Eq. \eqref{eq:EPtranpolPxyCPV}.
The interference between electro- (or magnetic-) and $CP$ violation terms $H_T$ contributes to non-diagonal correlations, thereby inducing a difference between $C_{ij}$ and $C_{ji}$ ($i \neq j, i, j =x, y, z$).
The electron mass correction introduces new angular distributions (those contan $\Sigma_T/\tau_e$ term) in $C_{xz}$ and $C_{yz}$ when at least one of the beams is transversely polarized.

Clearly above calculations confirm that the spin of final baryon and antibaryon are correlated in the $xz$ components even the initial electron beams are unpolarized in the case of $P$ and $CP$ conservation \cite{Faldt:2017kgy}.
A measurement of $C_{xz}$ allows for determine the real part of interference between electro- and magnetic form factors with unpolarized beams.
The interference of $P$ and $CP$ violation terms contributes to $C_{xz}$ and $C_{zx}$, and provokes difference between $C_{xz}$ and $C_{zx}$.
The $P$ and $CP$ violation separately induce non-zero double-spin polarization observables in the $xy$ and $yz$ components through different interference pattern with EMFFs.

Under the case of $P$ and $CP$ conservation, the identity $C_{xx} + C_{yy} + C_{zz} = 1$ holds. 
The $CP$ violation term breaks this identity to be
\bea
&& C_{xx} + C_{yy} + C_{zz} - 1 = \nn \\ && -\frac{4(\tau - 1) \left( I_{L}^-\sin^2 \theta + \frac{1}{\tau_e} I_{L}^+\cos^2 \theta \right)}{D}|H_T|^2 \,, \nn
\eea
if electron and positron beams are both longitudinally polarized, and
\bea
&& C_{xx} + C_{yy} + C_{zz} - 1 = \nn \\ && -\frac{4(\tau - 1) \left( (1+P_{T}^2\cos2\phi)\sin^2 \theta + \frac{1}{\tau_e} I_{T}^+\cos^2 \theta \right)}{D}|H_T|^2 \,, \nn
\eea
in the case of transversely polarized beams.

\section{Spin density matrix} \label{correlationmatrix}

The spin density matrix $C_{\mu \nu}$ of baryon-antibaryon pair are explicitly written down in a compact way because it is convenient to calculate the whole angular distributions of sequential decays in the helicity formalism \cite{Chen:2007zzf,Perotti:2018wxm,Salone:2022lpt,Fu:2023ose,Cao:2024tvz} and the quantum entanglement of hyperon-anti-hyperon pairs \cite{Tornqvist:1980af,Qian:2020ini,Fabbrichesi:2024rec,Wu:2024mtj,Hong:2025drg,Zhang:2026nwm,Li:2026bkf}.
By combining the single and double polarization observables in Sec.~\ref{singlespinpol} and Section~\ref{doublespinpol}, one has
\be
C_{\mu \nu} = \left(
\begin{array}{cccc}
 C_{00} & P_x & P_y    & P_z \\
 \bar{P}_{x} & C_{xx} & C_{xy} &  C_{xz} \\
 \bar{P}_{y} & C_{yx} & C_{yy} & C_{yz} \\
 \bar{P}_{z} & C_{zx} & C_{zy} & C_{zz}  \\
\end{array}
\right) \,,\label{eqn:cmatrixlong}
\ee
with $C_{00}$ being the normalized unpolarized cross section and $\bar{P}_{x,y,z} = P_{x,y,z} (\bar{Y})$.
The relative phases are defined as:
\begin{equation}
\begin{aligned}
\Delta\Phi_{EM} &= \arg(G_E/G_M) \,, \qquad
\Delta\Phi_{AT} &= \arg(F_A/H_T) \,, \\
\Delta\Phi_{AE} &= \arg(F_A/G_E) \,, \qquad
\Delta\Phi_{TM} &= \arg(H_T/G_M) \,, \\
\Delta\Phi_{AM} &= \arg(F_A/G_M) \,, \qquad
\Delta\Phi_{TE} &= \arg(H_T/G_E), \nn
\end{aligned}
\end{equation}
which correspond to the relative phases between the electric and magnetic form factors, $P$- and $CP$ violating strengths, $P$ violating strength and electric form factor, $CP$ violating strength and magnetic form factor, $P$-violating strength and magnetic form factor, and $CP$ violating strength and electric form factor, respectively. 
Note that only three relative phases among six are independent.

For the case of longitudinally polarized beams, $C_{\mu\nu}$ is decomposed into 17 matrices as follows:
\begin{widetext}
\begin{equation}
\begin{aligned}
C_{\mu\nu}&=I_L^-\biggl(C_{\mu\nu}(EM)+C_{\mu\nu}(AT)+C_{\mu\nu}(AE)+C_{\mu\nu}(TM)+C_{\mu\nu}(AM)+C_{\mu\nu}(TE)\biggl)\\
&+\Delta L \biggl(C_{\mu\nu}^{(L)}(EM)+C_{\mu\nu}^{(L)}(AT)+C_{\mu\nu}^{(L)}(AE)+C_{\mu\nu}^{(L)}(TM)+C_{\mu\nu}^{(L)}(AM)\biggr)\\
&+\frac{I_L^+}{\tau_e}\biggl(C_{\mu\nu}^{(m)}(EM)+C_{\mu\nu}^{(m)}(AT)+C_{\mu\nu}^{(m)}(AE)+C_{\mu\nu}^{(m)}(TM)+C_{\mu\nu}^{(m)}(AM)+C_{\mu\nu}^{(m)}(TE)\biggr) \,,
\label{eq:CmunvLongFinalEP}
\end{aligned}
\end{equation}
\end{widetext}
The letters in parentheses denote the involved form factors and the superscript $m$ denotes electron mass correction terms.
These matrices are categorized according to distinct combinations of the electric, magnetic, $P$-violating and $CP$-violating form factors and  unambiguously given as follows:
\begin{widetext}
\allowdisplaybreaks
\begin{align}
C_{\mu\nu}(EM) &= \xi_{EM} \times 
\begin{pmatrix}
1 + \alpha_{EM} \cos^2\theta & 0 & \beta_{EM} \sin\theta \cos\theta & 0 \\
0 & \sin^2\theta & 0 & -\gamma_{EM} \sin\theta \cos\theta \\
\beta_{EM} \sin\theta \cos\theta & 0 & -\alpha_{EM} \sin^2\theta & 0 \\
0 & -\gamma_{EM} \sin\theta \cos\theta & 0 & \alpha_{EM} + \cos^2\theta
\end{pmatrix},
\\
C_{\mu\nu}(AT) &= \xi_{AT} (\tau - 1) \times 
\begin{pmatrix}
1 - \alpha_{AT} \cos^2\theta & 0 & \gamma_{AT} \sin\theta \cos\theta & 0 \\
0 & -\sin^2\theta & 0 & \beta_{AT} \sin\theta \cos\theta \\
-\gamma_{AT} \sin\theta \cos\theta & 0 & -\alpha_{AT} \sin^2\theta & 0 \\
0 & -\beta_{AT} \sin\theta \cos\theta & 0 & -\alpha_{AT} + \cos^2\theta
\end{pmatrix},\label{cmunvAT}
\\
C_{\mu\nu}(AE) &= \xi_{AE} \frac{\sqrt{\tau - 1}}{\sqrt{\tau}} \times 
\begin{pmatrix}
0 & \gamma_{AE} \sin\theta \cos\theta & 0 & 0 \\
\gamma_{AE} \sin\theta \cos\theta & 0 & 0 & 0 \\
0 & 0 & 0 & \beta_{AE} \sin\theta \cos\theta \\
0 & 0 & \beta_{AE} \sin\theta \cos\theta & 0
\end{pmatrix},
\\
C_{\mu\nu}(TM) &= \xi_{TM} \sqrt{\tau (\tau - 1)} \times 
\begin{pmatrix}
0 & -\beta_{TM} \sin\theta \cos\theta & 0 & 0 \\
\beta_{TM} \sin\theta \cos\theta & 0 & 0 & 0 \\
0 & 0 & 0 & \gamma_{TM} \sin\theta \cos\theta \\
0 & 0 & -\gamma_{TM} \sin\theta \cos\theta & 0
\end{pmatrix},\label{cmunvTM}
\\
C_{\mu\nu}(AM) &= -\xi_{AM} \sqrt{\tau - 1} \times 
\begin{pmatrix}
0 & 0 & 0 & \gamma_{AM} (1 + \cos^2\theta) \\
0 & 0 & \beta_{AM} \sin^2\theta & 0 \\
0 & \beta_{AM} \sin^2\theta & 0 & 0 \\
\gamma_{AM} (1 + \cos^2\theta) & 0 & 0 & 0
\end{pmatrix},
\\
C_{\mu\nu}(TE) &= \xi_{TE} \sqrt{\tau - 1} \times 
\begin{pmatrix}
0 & 0 & 0 & -\beta_{TE} \sin^2\theta \\
0 & 0 & \gamma_{TE} \sin^2\theta & 0 \\
0 & -\gamma_{TE} \sin^2\theta & 0 & 0 \\
\beta_{TE} \sin^2\theta & 0 & 0 & 0
\end{pmatrix},\label{cmunvTE}
\\
C_{\mu\nu}^{(L)}(EM) &= \xi_{EM} \times 
\begin{pmatrix}
0 & -\gamma_{EM}\sin\theta & 0 & (1+\alpha_{EM}\cos\theta) \\
-\gamma_{EM}\sin\theta & 0 & 0 & 0 \\
0 & 0 & 0 & \beta_{EM}\sin\theta \\
(1+\alpha_{EM}\cos\theta) & 0 & \beta_{EM}\sin\theta & 0
\end{pmatrix},
\\
C_{\mu\nu}^{(L)}(AT) &= \xi_{AT}(\tau-1) \times 
\begin{pmatrix}
0 & -\beta_{AT}\sin\theta & 0 & (1-\alpha_{AT})\cos\theta \\
\beta_{AT}\sin\theta & 0 & 0 & 0 \\
0 & 0 & 0 & -\gamma_{AT}\sin\theta \\
(1-\alpha_{AT})\cos\theta & 0 & \gamma_{AT}\sin\theta & 0
\end{pmatrix}, \label{cmunvLAT}
\\
C_{\mu\nu}^{(L)}(AE) &= \xi_{AE}\frac{\sqrt{\tau-1}}{\sqrt{\tau}} \times 
\begin{pmatrix}
0 & 0 & \beta_{AE}\sin\theta & 0 \\
0 & 0 & 0 & \gamma_{AE}\sin\theta \\
\beta_{AE}\sin\theta & 0 & 0 & 0 \\
0 & \gamma_{AE}\sin\theta & 0 & 0
\end{pmatrix},
\\
C_{\mu\nu}^{(L)}(TM) &= \xi_{TM}\sqrt{\tau(\tau-1)} \times 
\begin{pmatrix}
0 & 0 & -\gamma_{TM}\sin\theta & 0 \\
0 & 0 & 0 & \beta_{TM}\sin\theta \\
\gamma_{TM}\sin\theta & 0 & 0 & 0 \\
0 & -\beta_{TM}\sin\theta & 0 & 0
\end{pmatrix}, \label{cmunvLTM}
\\
C_{\mu\nu}^{(L)}(AM) &= 2\xi_{AM}\sqrt{\tau-1} \times 
\begin{pmatrix}
\gamma_{AM}\cos\theta & 0 & 0 & 0 \\
0 & 0 & 0 & 0 \\
0 & 0 & 0 & 0 \\
0 & 0 & 0 & \gamma_{AM}\cos\theta
\end{pmatrix},
\\
C_{\mu\nu}^{(m)}(EM) &= \frac{1}{2}\xi_{EM} \times 
\begin{pmatrix}
1-\alpha_{EM}\cos2\theta & 0 & -\beta_{EM} \sin2\theta & 0 \\
0 & \cos2\theta-\alpha_{EM} & 0 & \gamma_{EM} \sin2\theta \\
-\beta_{EM} \sin2\theta & 0 & 1-\alpha_{EM}\cos2\theta & 0 \\
0 & \gamma_{EM} \sin2\theta & 0 & \alpha_{EM}-\cos2\theta
\end{pmatrix},
\\
C_{\mu\nu}^{(m)}(AT) &= \frac{1}{2}\xi_{AT} (\tau - 1) \times 
\begin{pmatrix}
1+\alpha_{AT}\cos2\theta & 0 & -\gamma_{AT} \sin2\theta & 0 \\
0 & -\alpha_{AT}-\cos2\theta & 0 & -\beta_{AT} \sin2\theta \\
\gamma_{AT} \sin2\theta & 0 & 2\alpha_{AT}+\alpha_{AT}\cos2\theta-1 & 0 \\
0 & \beta_{AT} \sin2\theta & 0 & -\alpha_{AT}-\cos2\theta
\end{pmatrix},\label{cmunvmassAT}
\\
C_{\mu\nu}^{(m)}(AM) &= \xi_{AM} \sqrt{\tau - 1} \times 
\begin{pmatrix}
0 & 0 & 0 & -\gamma_{AM} \sin^2\theta \\
0 & 0 & \beta_{AM} \sin^2\theta & 0 \\
0 & \beta_{AM} \sin^2\theta & 0 & 0 \\
-\gamma_{AM} \sin^2\theta & 0 & 0 & 0
\end{pmatrix},
\\
C_{\mu\nu}^{(m)}(TE) &= \xi_{TE} \sqrt{\tau - 1} \times 
\begin{pmatrix}
0 & 0 & 0 & -\beta_{TE} \cos^2\theta \\
0 & 0 & \gamma_{TE} \cos^2\theta & 0 \\
0 & -\gamma_{TE} \cos^2\theta & 0 & 0 \\
\beta_{TE} \cos^2\theta & 0 & 0 & 0
\end{pmatrix} \,,\label{cmunvmassTE}
\end{align}
\end{widetext}
and $C_{\mu\nu}^{(m)}(AE) =- C_{\mu\nu}(AE)$, $C_{\mu\nu}^{(m)}(TM) =- C_{\mu\nu}(TM)$.
Here $\alpha$, $\beta$ and $\gamma$ as listed in Appendix~\ref{apdx:parameters} are introduced for convenience, and they are similar to  hyperon decay parameters first proposed by Lee and Yang for parity violation decay \cite{Lee:1956qn}.
Note that only four among six $\xi$ variables are independent.
For the same group of $\alpha$, $\beta$ and $\gamma$, $\alpha^2 + \beta^2 + \gamma^2 =1$ is fulfilled.
For the case of transversely polarized beams, $C_{\mu\nu}$ is decomposed into 23 matrices as follows:
\begin{widetext}
\begin{equation}
\begin{aligned}
C_{\mu\nu}&=C_{\mu\nu}(EM)+C_{\mu\nu}(AT)+C_{\mu\nu}(AE)+C_{\mu\nu}(TM)+C_{\mu\nu}(AM)+C_{\mu\nu}(TE)\\
&+P_{T}^2\biggl(C_{\mu\nu}^{(T)}(EM)+C_{\mu\nu}^{(T)}(AT)+C_{\mu\nu}^{(T)}(AE)+C_{\mu\nu}^{(T)}(AM)+C_{\mu\nu}^{(T)}(TM)+C_{\mu\nu}^{(T)}(TE)\biggr)\\
&+\frac{I_{T}^{+}}{\tau_e}\biggl(C_{\mu\nu}^{(m)}(EM)+C_{\mu\nu}^{(m)}(AT)+C_{\mu\nu}^{(m)}(AE)+C_{\mu\nu}^{(m)}(TM)+C_{\mu\nu}^{(m)}(AM)+C_{\mu\nu}^{(m)}(TE)\biggr)\\
&+\frac{\Sigma_T}{\sqrt{\tau_e}}\biggl(C_{\mu\nu}^{(T_m)}(EM)+C_{\mu\nu}^{(T_m)}(AT)+C_{\mu\nu}^{(T_m)}(AE)+C_{\mu\nu}^{(T_m)}(AM)+C_{\mu\nu}^{(T_m)}(TM)\biggl)\,,
\label{eq:CmunvTranFinalEP}
\end{aligned}
\end{equation}
\end{widetext}
where the six matrices $C_{\mu\nu}$ and six matrices $C_{\mu\nu}^{(m)}$ are the same as those in the above case of unpolarized and longitudinally polarized beams.
Other matrices are
\begin{widetext}
\allowdisplaybreaks
\begin{align}
C_{\mu\nu}^{(T)}(EM) &= \xi_{EM} \nonumber\\
&\times 
\begin{pmatrix}
-\alpha_{EM}\sin^2\theta\cos2\phi & \beta_{EM}\sin\theta\sin2\phi & \beta_{EM}\sin\theta\cos\theta\cos2\phi & 0 \\
\beta_{EM}\sin\theta\sin2\phi & -(\alpha_{EM}+\cos^2\theta)\cos2\phi & (1+\alpha_{EM})\cos\theta\sin2\phi & -\gamma_{EM}\sin\theta\cos\theta\cos2\phi \\
\beta_{EM}\sin\theta\cos\theta\cos2\phi & (1+\alpha_{EM})\cos\theta\sin2\phi & (1+\alpha_{EM}\cos^2\theta)\cos2\phi & \gamma_{EM}\sin\theta\sin2\phi \\
0 & -\gamma_{EM}\sin\theta\cos\theta\cos2\phi & \gamma_{EM}\sin\theta\sin2\phi & -\sin^2\theta\cos2\phi
\end{pmatrix},
\\
C_{\mu\nu}^{(T)}(AT) &= \xi_{AT} (\tau - 1) \nonumber\\
&\times 
\begin{pmatrix}
\alpha_{AT}\sin^2\theta\cos2\phi & \gamma_{AT}\sin\theta\sin2\phi & \gamma_{AT}\sin\theta\cos\theta\cos2\phi & 0 \\
-\gamma_{AT}\sin\theta\sin2\phi & (\cos^2\theta-\alpha_{AT})\cos2\phi & (\alpha_{AT}-1)\cos\theta\sin2\phi & \beta_{AT}\sin\theta\cos\theta\cos2\phi \\
-\gamma_{AT}\sin\theta\cos\theta\cos2\phi & (\alpha_{AT}-1)\cos\theta\sin2\phi & (\alpha_{AT}\cos^2\theta-1)\cos2\phi & -\beta_{AT}\sin\theta\sin2\phi \\
0 & -\beta_{AT}\sin\theta\cos\theta\cos2\phi & \beta_{AT}\sin\theta\sin2\phi & -\sin^2\theta\cos2\phi
\end{pmatrix}, \label{TcmunvAT}
\\
C_{\mu\nu}^{(T)}(AE) &= \xi_{AE} \frac{\sqrt{\tau - 1}}{\sqrt{\tau}} \nonumber\\
&\times 
\begin{pmatrix}
0 & \gamma_{AE}\sin\theta\cos\theta\cos2\phi & -\gamma_{AE}\sin\theta\sin2\phi & 0 \\
\gamma_{AE}\sin\theta\cos\theta\cos2\phi & 0 & 0 & \beta_{AE}\sin\theta\sin2\phi \\
-\gamma_{AE}\sin\theta\sin2\phi & 0 & 0 & \beta_{AE}\sin\theta\cos\theta\cos2\phi\\
0 & \beta_{AE}\sin\theta\sin2\phi & \beta_{AE}\sin\theta\cos\theta\cos2\phi & 0
\end{pmatrix},
\\
C_{\mu\nu}^{(T)}(AM) &= \xi_{AM} \sqrt{\tau - 1} \nonumber\\
&\times 
\begin{pmatrix}
0 & 0 & 0 & \gamma_{AM}\sin^2\theta\cos2\phi \\
0 & 2\beta_{AM}\cos\theta\sin2\phi & \beta_{AM}(1+\cos^2\theta)\cos2\phi & 0 \\
0 & \beta_{AM}(1+\cos^2\theta)\cos2\phi & -2\beta_{AM}\cos\theta\sin2\phi & 0 \\
\gamma_{AM}\sin^2\theta\cos2\phi & 0 & 0 & 0
\end{pmatrix},
\\
C_{\mu\nu}^{(T)}(TM) &= \xi_{TM} \sqrt{\tau (\tau - 1)} \nonumber\\
&\times 
\begin{pmatrix}
0 & -\beta_{TM}\sin\theta\cos\theta\cos2\phi & \beta_{TM}\sin\theta\sin2\phi & 0 \\
\beta_{TM}\sin\theta\cos\theta\cos2\phi & 0 & 0 & \gamma_{TM}\sin\theta\sin2\phi \\
-\beta_{TM}\sin\theta\sin2\phi & 0 & 0 & \gamma_{TM}\sin\theta\cos\theta\cos2\phi \\
0 & -\gamma_{TM}\sin\theta\sin2\phi & -\gamma_{TM}\sin\theta\cos\theta\cos2\phi & 0
\end{pmatrix}, \label{TcmunvTM}
\\
C_{\mu\nu}^{(T)}(TE) &= \xi_{TE} \sqrt{\tau - 1} \times 
\begin{pmatrix}
0 & 0 & 0 & -\beta_{TE}\sin^2\theta\cos2\phi \\
0 & 0 & \gamma_{TE}\sin^2\theta\cos2\phi & 0 \\
0 & -\gamma_{TE}\sin^2\theta\cos2\phi & 0 & 0 \\
\beta_{TE}\sin^2\theta\cos2\phi & 0 & 0 & 0
\end{pmatrix}, \label{TcmunvTE}
\\
C_{\mu\nu}^{(T_m)}(EM) &= \xi_{EM} \times 
\begin{pmatrix}
0 & \gamma_{EM}\cos\theta\sin\phi & \gamma_{EM}\cos\phi & (1+\alpha_{EM})\sin\theta\sin\phi \\
\gamma_{EM}\cos\theta\sin\phi & 0 & 0 & \beta_{EM}\cos\phi \\
\gamma_{EM}\cos\phi & 0 & 0 & -\beta_{EM}\cos\theta\sin\phi \\
(1+\alpha_{EM})\sin\theta\sin\phi & \beta_{EM}\cos\phi & -\beta_{EM}\cos\theta\sin\phi & 0
\end{pmatrix},
\\
C_{\mu\nu}^{(T_m)}(AT) &= \xi_{AT} (\tau - 1) \times 
\begin{pmatrix}
0 & \beta_{AT}\cos\theta\sin\phi & \beta_{AT}\cos\phi & (1-\alpha_{AT})\sin\theta\sin\phi \\
-\beta_{AT}\cos\theta\sin\phi & 0 & 0 & -\gamma_{AT}\cos\phi \\
-\beta_{AT}\cos\phi & 0 & 0 & \gamma_{AT}\cos\theta\sin\phi \\
(1-\alpha_{AT})\sin\theta\sin\phi & \gamma_{AT}\cos\phi & -\gamma_{AT}\cos\theta\sin\phi & 0
\end{pmatrix}, \label{TmcmunvAT}
\\
C_{\mu\nu}^{(T_m)}(AE) &= \xi_{AE} \frac{\sqrt{\tau - 1}}{\sqrt{\tau}} \times 
\begin{pmatrix}
0 & \beta_{AE}\cos\phi & -\beta_{AE}\cos\theta\sin\phi & 0 \\
\beta_{AE}\cos\phi & 0 & 0 & -\gamma_{AE}\cos\theta\sin\phi \\
-\beta_{AE}\cos\theta\sin\phi & 0 & 0 & -\gamma_{AE}\cos\phi\\
0 & -\gamma_{AE}\cos\theta\sin\phi & -\gamma_{AE}\cos\phi & 0
\end{pmatrix},
\\
C_{\mu\nu}^{(T_m)}(AM) &= \xi_{AM} \sqrt{\tau - 1} \times 
\begin{pmatrix}
-2\gamma_{AM}\sin\theta\sin\phi & 0 & 0 & 0 \\
0 & 0 & 0 & 0 \\
0 & 0 & 0 & 0 \\
0 & 0 & 0 & -2\gamma_{AM}\sin\theta\sin\phi
\end{pmatrix},
\\
C_{\mu\nu}^{(T_m)}(TM) &= \xi_{TM} \sqrt{\tau (\tau - 1)} \times 
\begin{pmatrix}
0 & -\gamma_{TM}\cos\phi & \gamma_{TM}\cos\theta\sin\phi & 0 \\
\gamma_{TM}\cos\phi & 0 & 0 & -\beta_{TM}\cos\theta\sin\phi \\
-\gamma_{TM}\cos\theta\sin\phi & 0 & 0 & -\beta_{TM}\cos\phi \\
0 & \beta_{TM}\cos\theta\sin\phi & \beta_{TM}\cos\phi & 0
\end{pmatrix}\, \label{TmcmunvTM}
\end{align}
\end{widetext}

Above results reveal that the matrices containing $CP$-violating form factors, see Eqs. \eqref{cmunvAT}, \eqref{cmunvTM}, \eqref{cmunvTE}, \eqref{cmunvLAT}, \eqref{cmunvLTM}, \eqref{cmunvmassAT}, \eqref{cmunvmassTE}, \eqref{TcmunvAT}, \eqref{TcmunvTM}, \eqref{TcmunvTE},  \eqref{TmcmunvAT}, and \eqref{TmcmunvTM} are antisymmetric, and the remaining matrices are all symmetric. 
It can be seen that moduli of the form factors and relative phases between them are what can be determined through a measurement of full angular distribution.
The recent BESIII measurement of $e^+e^- \to J/\psi \to \Lambda (p \pi^-) \bar\Lambda (\overline p \pi^+)$ took one of the electro-magnetic form factors as a real number~\cite{BESIII:2025vxm}.
In Table~\ref{tab:Parameters}, an alternative prescription of parameters based on our analysis is given by neglecting the electron mass.
The numerical values are deduced from the electromagnetic, $P$ and $CP$ violating form factors at $J/\psi$ energy by BESIII~\cite{BESIII:2025vxm} and the uncertainties are calculated through naive error propagation so the electric and magnetic form factors shall be compared to previous BESIII results with caution~\cite{BESIII:2022qax,BESIII:2018cnd}.
Note that only seven parameters in Table~\ref{tab:Parameters} are independent, for instance one
can choose four $\xi$ and three relative phases.
The current precision for $\Lambda$ at $J/\psi$ energy are
\be
\lf|\frac{F_A}{G_M} \rg| =0.0051 \pm 0.0027 \,, \lf|\frac{H_T}{G_M} \rg| =0.0026 \pm 0.0019 \,,
\ee
so the electron mass correction is not important at present.
For the proposed Super Tau-Charm Facility \cite{Achasov:2023gey}, the sensitivity is estimated to be two orders higher \cite{Ovsiannikov:2025gcy} so electron mass correction is recommended to be considered.

\begin{table}[htbp]
\centering
\caption{Numerical parameters with uncertainties for $e^+e^- \to J/\psi \to \Lambda (p \pi^-) \bar\Lambda (\overline p \pi^+)$ deduced from the BESIII  result~\cite{BESIII:2025vxm}. The relative phases $\Delta\Phi$ are given in unit of rad.}
\label{tab:Parameters}
\begin{tabular}{lc}
\hline
Parameter & Value\\
\hline
$\xi_{EM}$ & $(3.5 \pm 0.13) \times 10^{-6}$ \\
$\xi_{AT}$ & $(5.3 \pm 4.5) \times 10^{-11}$ \\
$\xi_{AE}$ & $(1.778 \pm 0.064) \times 10^{-6}$ \\
$\xi_{TM}$ & $(2.59 \pm 0.13) \times 10^{-6}$ \\
$\xi_{AM}$ & $(2.59 \pm 0.13) \times 10^{-6}$ \\
$\xi_{TE}$ & $(1.778 \pm 0.064) \times 10^{-6}$ \\
$\Delta\Phi_{EM}$ & $0.755 \pm 0.018 $ \footnote{$0.7521 \pm 0.0078 $~\cite{BESIII:2022qax} and $0.740 \pm 0.013 $\cite{BESIII:2018cnd}} \\
$\Delta\Phi_{AT}$ & $-4.26 \pm 0.82 $ \\
$\Delta\Phi_{AE}$ & $-2.62 \pm 0.43 $ \\
$\Delta\Phi_{TM}$ & $2.39 \pm 0.70 $ \\
$\Delta\Phi_{AM}$ & $-1.87 \pm 0.43 $ \\
$\Delta\Phi_{TE}$ & $1.64 \pm 0.70 $ \\
$\alpha_{EM}$ & $0.475 \pm 0.019$ \footnote{$0.4748 \pm 0.0038$~\cite{BESIII:2022qax} and $0.461 \pm 0.009$~\cite{BESIII:2018cnd} } \\
$\alpha_{AT}$ & $-0.32 \pm 1.23$ \\
$\alpha_{AE}$ & $0.999960 \pm 0.000030$ \\
$\alpha_{TM}$ & $0.999993 \pm 0.0000072$ \\
$\alpha_{AM}$ & $0.999973 \pm 0.000021$ \\
$\alpha_{TE}$ & $0.999989 \pm 0.000011$ \\
$\beta_{EM}$ & $0.603 \pm 0.013$ \\
$\beta_{AT}$ & $0.85 \pm 0.51$ \\
$\beta_{AE}$ & $-0.0044 \pm 0.0038$ \\
$\beta_{TM}$ & $0.0026 \pm 0.0023$ \\
$\beta_{AM}$ & $-0.0071 \pm 0.0028$ \\
$\beta_{TE}$ & $0.0046 \pm 0.0023$ \\
$\gamma_{EM}$ & $0.641 \pm 0.013$ \\
$\gamma_{AT}$ & $-0.41 \pm 0.72$ \\
$\gamma_{AE}$ & $-0.0077 \pm 0.0035$ \\
$\gamma_{TM}$ & $-0.0028 \pm 0.0023$ \\
$\gamma_{AM}$ & $-0.0021 \pm 0.0032$ \\
$\gamma_{TE}$ & $-0.0003 \pm 0.0032$ \\
\hline
\end{tabular}
\end{table}

\section{Summary} \label{conclusion}

This paper explicitly gives the spin density matrix  for spin half baryon-antibaryon pairs produced through one-photon exchange in polarized electron-positron annihilation.
The $P$ and $CP$ violation in the vertex of virtual photon coupling to the baryon pairs are considered and the electron mass correction is included.
The calculation of polarization and correlation is performed by the method of spin projection operator and the final results are obtained in the centre-of-mass frame. 
The whole spin density matrix is written down in a compact way for later calculation of the full angular distributions of sequential decays within the helicity formalism and and the quantum entanglement of hyperon-anti-hyperon pairs.
The beam polarization introduces new $\sin\theta$ and $\cos\theta$ dependence to observables, confronting the angular distribution of TPE contribution.
The transverse polarization of beams further induces azimuthal angular $\phi$ dependence.

Our results are consistent with those for the unpolarized electron and positron beams and portable to the experimental analysis of longitudinal and transversely polarized beams \cite{Zhang:2025oks,Salone:2022lpt,Cao:2024tvz}. 
The parameters of $e^+e^- \to J/\psi \to \Lambda (p \pi^-) \bar\Lambda (\overline p \pi^+)$ in our prescription are infered from recent BESIII data.
Our framework would be also used to study the $P$ and $CP$ violation in $\tau$ lepton pair production~\cite{Banerjee:2022sgf,Huang:2025ghw}.
Even though the electron mass corrections are minor, properly accounting for them is crucial to avoid misleading interpretations when probing similarly delicate $P$ and $CP$-violating signals.

\bigskip

\begin{acknowledgments}

We are grateful to Yong Du for useful discussions. This work is supported by the National Key R\&D Program of China under Grant No. 2023YFA1606703, and the National Natural Science Foundation of China (Grants Nos. 12547111 and 12165022), and Yunnan Fundamental Research Project under Contract No. 202301AT070162.

\end{acknowledgments}

\bigskip

\appendix

\section{Definition of parameters}\label{apdx:parameters}
The parameters in spin density matrix are listed here:
\begin{equation}
\begin{split}
\xi_{EM} &= \frac{1}{\tau} (\tau |G_M|^2 + |G_E|^2) \,, \quad
\alpha_{EM} = \frac{\tau |G_M|^2 - |G_E|^2}{\tau |G_M|^2 + |G_E|^2} \,, \\
\beta_{EM} &= \sqrt{1 - \alpha_{EM}^2} \sin(\Delta\Phi_{EM}) \,, \\
\gamma_{EM} &= \sqrt{1 - \alpha_{EM}^2} \cos(\Delta\Phi_{EM}) \,,
\end{split}
\end{equation}
for the usual parameters related to electric and magnetic form factors.
Similarly the parameters for other form factors are defined as 
\begin{equation}
\begin{aligned}
\xi_{AT} &= \frac{1}{\tau} (\tau |H_T|^2 + |F_A|^2) \,, \quad
\alpha_{AT} = \frac{\tau |H_T|^2 - |F_A|^2}{\tau |H_T|^2 + |F_A|^2} \,, \\
\beta_{AT} &= \sqrt{1 - \alpha_{AT}^2} \sin(\Delta\Phi_{AT}) \,, \\
\gamma_{AT} &= \sqrt{1 - \alpha_{AT}^2} \cos(\Delta\Phi_{AT}) \,,
\end{aligned}
\end{equation}
\begin{equation}
\begin{aligned}
\xi_{AE} &= \frac{1}{\tau} (\tau |G_E|^2 + |F_A|^2) \,, \quad
\alpha_{AE} = \frac{\tau |G_E|^2 - |F_A|^2}{\tau |G_E|^2 + |F_A|^2} \,, \\
\beta_{AE} &= \sqrt{1 - \alpha_{AE}^2} \sin(\Delta\Phi_{AE}) \,, \\
\gamma_{AE} &= \sqrt{1 - \alpha_{AE}^2} \cos(\Delta\Phi_{AE}) \,,
\end{aligned}
\end{equation}
\begin{equation}
\begin{aligned}
\xi_{TM} &= \frac{1}{\tau} (\tau |G_M|^2 + |H_T|^2) \,, \quad
\alpha_{TM} = \frac{\tau |G_M|^2 - |H_T|^2}{\tau |G_M|^2 + |H_T|^2} \,, \\
\beta_{TM} &= \sqrt{1 - \alpha_{TM}^2} \sin(\Delta\Phi_{TM}) \,, \\
\gamma_{TM} &= \sqrt{1 - \alpha_{TM}^2} \cos(\Delta\Phi_{TM}) \,,
\end{aligned}
\end{equation}
\begin{equation}
\begin{aligned}
\xi_{AM} &= \frac{1}{\tau} (\tau |G_M|^2 + |F_A|^2) \,,\quad
\alpha_{AM} = \frac{\tau |G_M|^2 - |F_A|^2}{\tau |G_M|^2 + |F_A|^2} \,,\\
\beta_{AM} &= \sqrt{1 - \alpha_{AM}^2} \sin(\Delta\Phi_{AM}) \,,\\
\gamma_{AM} &= \sqrt{1 - \alpha_{AM}^2} \cos(\Delta\Phi_{AM}) \,,
\end{aligned}
\end{equation}
\begin{equation}
\begin{aligned}
\xi_{TE} &= \frac{1}{\tau} (\tau |G_E|^2 + |H_T|^2) \,,\quad
\alpha_{TE} = \frac{\tau |G_E|^2 - |H_T|^2}{\tau |G_E|^2 + |H_T|^2} \,,\\
\beta_{TE} &= \sqrt{1 - \alpha_{TE}^2} \sin(\Delta\Phi_{TE}) \,,\\
\gamma_{TE} &= \sqrt{1 - \alpha_{TE}^2} \cos(\Delta\Phi_{TE}) \,,
\end{aligned}
\end{equation}
The corresponding numerical values of $e^+e^- \to J/\psi \to \Lambda (p \pi^-) \bar\Lambda (\overline p \pi^+)$ in Table~\ref{tab:Parameters}
are deduced from the BESIII measurement~\cite{BESIII:2025vxm}.

\section{The contraction of antisymmetric tensor}\label{apdx:asy}
The antisymmetric tensors employed in the computation of unpolarized beams case are explicitly enumerated herein, whereas around 150 non-zero components for the case of polarized beams are omitted for simplicity:
\bea
\epsilon^{k_1 k_2 p_1 s_{1y}} &=& -\frac{s^{3/2}}{4} \beta_e \beta_Y \sin\theta \nn \\
\epsilon^{k_1 k_2 p_1 s_{2y}} &=& -\frac{s^{3/2}}{4} \beta_e \beta_Y \sin\theta \nn \\
\epsilon^{k_1 k_2 s_{1x} s_{2y}} &=& -\frac{s}{2} \beta_e \cos\theta \nn \\
\epsilon^{k_1 p_1 s_{1x} s_{2y}} &=& \frac{s}{4} \left( \beta_Y - \beta_e \cos\theta \right) \nn \\
\epsilon^{k_2 p_1 s_{1x} s_{2y}} &=& \frac{s}{4} \left( \beta_Y + \beta_e \cos\theta \right) \nn \\
\epsilon^{k_1 k_2 s_{1y} s_{2x}} &=& \frac{s}{2} \beta_e \cos\theta \nn \\
\epsilon^{k_1 p_1 s_{1y} s_{2x}} &=& \frac{s}{4} \left( \beta_Y - \beta_e \cos\theta \right) \nn \\
\epsilon^{k_2 p_1 s_{1y} s_{2x}} &=& \frac{s}{4} \left( \beta_Y + \beta_e \cos\theta \right) \nn \\
\epsilon^{k_1 k_2 s_{1y} s_{2z}} &=& \frac{s}{2} \sqrt{\tau} \beta_e \sin\theta \nn \\
\epsilon^{k_1 p_1 s_{1y} s_{2z}} &=& \frac{s}{4} \sqrt{\tau} \left(2 - \frac{1}{\tau}\right) \nn \\
\epsilon^{k_2 p_1 s_{1y} s_{2z}} &=& -\frac{s}{4} \sqrt{\tau} \left(2 - \frac{1}{\tau}\right) \nn \\
\epsilon^{k_1 k_2 s_{1z} s_{2y}} &=& -\frac{s}{2} \sqrt{\tau} \beta_e \sin\theta \nn \\
\epsilon^{k_1 p_1 s_{1z} s_{2y}} &=& -M^2 \sqrt{\tau} \beta_e \sin\theta \nn \\
\epsilon^{k_2 p_1 s_{1z} s_{2y}} &=& M^2 \sqrt{\tau} \beta_e \sin\theta \nn
\eea
with $\beta_e=\sqrt{1 - \frac{4 m^2}{s}},\beta_Y=\sqrt{1 - \frac{4 M^2}{s}}$.

\bibliography{NEFF_all.bib}

\end{document}